\documentclass[runningheads]{llncs}
\usepackage{graphicx}
\usepackage{amsmath}

\begin{document}
\title{Deep Subspace analysing for Semi-Supervised multi-label classification of Diabetic Foot Ulcer}
%
%
\author{Azadeh Alavi\inst{1,2}}
\institute{AI Discipline, School of Computing Technologies, RMIT University, VIC 3001, Australia 
\email{azadeh.alavi@rmit.edu.au} \and Bioinformatics lab, Baker Heart and Diabetes Institute, VIC 3002, Australia}
%
%
%
\maketitle              
\begin{abstract}
 Diabetes is a global raising pandemic. Diabetes patients are at risk of developing foot ulcer that usually leads to limb amputation. In order to develop a self monitoring mobile application, in this work, we propose a novel deep subspace analysis pipeline for semi-supervised diabetic foot ulcer mulit-label classification. To avoid any chance of over-fitting, unlike recent state of the art deep semi-supervised methods, the proposed pipeline dose not include any data augmentation. Whereas, after extracting deep features, in order to make the representation shift invariant, we employ variety of data augmentation methods on each image and generate an image-sets, which is then mapped into a linear subspace. Moreover, the proposed pipeline reduces the cost of retraining when more new unlabelled data become available. Thus, the first stage of the pipeline employs the concept of transfer learning for feature extraction purpose through modifying and retraining a deep convolutional network architect known as Xception. Then, the output of a mid-layer is extracted to generate an image set representer of any given image with help of data augmentation methods. At this stage, each image is transferred to a linear subspace which is a point on a Grassmann Manifold topological space. Hence, to perform analyse them, the geometry of such manifold must be considered. As such, each labelled image is represented as a vector of distances to number of unlabelled images using geodesic distance on Grassmann manifold. Finally, Random Forest is trained for multi-label classification of diabetic foot ulcer images. The method is then evaluated on the blind test set provided by DFU2021 competition, and the result considerable improvement compared to using classical transfer learning with data augmentation.

\end{abstract}
\section{Introduction}

Diabetes is a raising universal problem that affects 425 million people which is expected to rise to 629 million people by 2045 \cite{info,DFUC2021}. One in three diabetic patients are likely to develop Diabetic Foot Ulcers (DFU) which is a serious complication of diabetes, and can lead to limb amputation, or even death if it is with infection and ischaemia \cite{info2}. 
However, as diabetes patients can loose sensation in their foot, it is hard for them to identify the development of such ulcer. In an effort to develop a self monitoring technology, in this work we study the classification of such ulcers.

\begin{figure}
\begin{center}
\includegraphics[width=0.5\textwidth]{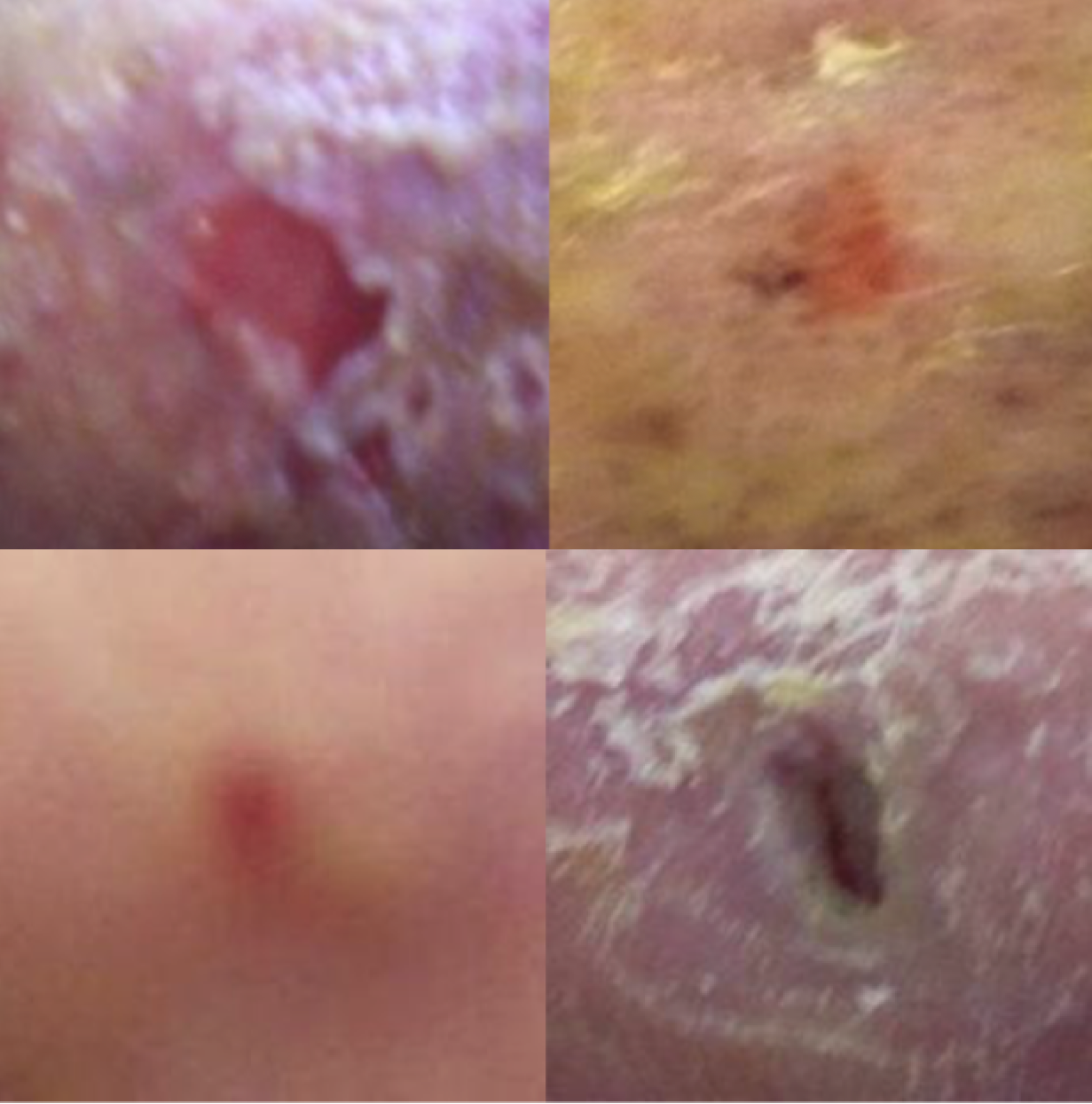}
\caption{An example of early stage DFU \cite{DFUC2021} } \label{fig3}
\end{center}
\end{figure}

Deep Learning based methods have achieved immense performance in variety of computer vision related tasks. However, their strong performance highly depends on the size of the presented training set. Preparing a sufficiently large labeled dataset can demand high labor cost. Hence, semi-supervised learning methods that take advantage of deep learning based models has become a topic of research interest in recent years ~\cite{ref_article2,ref_conf2,ref_article3}.

While earlier work limited the use of unlabeled data to pre-training stage ~\cite{ref_conf4,ref_conf5}, more recent researches studied exploiting unlabeled data in the entire training stage ~\cite{ref_conf4,ref_conf5}. \\
Recent advancements in semi-supervised deep learning methods are commonly gained by modifying the loss function, which can be achieved by adding regularization over the unlabelled data ~\cite{ref_article5}.
However, as new unlabeled data become available, all these methods must retrain the deep network in order to take advantage of such newly available data. That is both time and resource consuming.\\
In addition, the majority of the recent approaches aim to fuse the inputs into coherent clusters by adding noise and smoothing the mapping function locally. In cases when the raw data already contain noise, such approaches would no longer be relevant.

In this work, we propose a novel 3-stage deep semi-supervised pipeline for diabetic foot ulcer classification purpose. First, we employ the concept of transfer learning to generate a discriminate set of features. Then, to create a shift invariant representor of any given image, we use augmentation tools to create an image set for each given image. Next, we map such image sets into linear subspaces which are a point on Grassman manifold topological space. That is then followed by computing the geodesic distance between each labeled point to centroid of un-labled points using k-median algorithm. That would generate the final representor of each image as a vector of relationships. That enables the method to take advantage of unlabeled data. Finally, we use the computed features in stage.2 to train Xception \cite{Xception} network for classification purpose.\\

The {\bfseries contributions} of this work are summaries bellow:
\begin{itemize}
    \item The proposed method offers a novel pipeline for deep subspace analysis for semi-supervised multi-label classification.
    \item The proposed pipeline generates an shift invariant representor of an image by first generating an image set from each given image through employing augmentation tools. It then maps each image set to a linear subspace which is a point on a Grassmann Manifold. Then, it employs the geodesic distance to redefine each point as a vector of distance to number of unlabled images.
    \item In general, the likelihood of new unlabelled data becoming available is always much higher compared to that for labeled data, as labeling the data is very resource demanding. The proposed method is designed so the retraining of the algorithm with more unlabeled data would not require retraining the network; hence is not time and resource consuming. 
\end{itemize}
 
 We have evaluated our method on DFU2021 blind test set, and the result shows promising performance. The result proved that the proposed deep relation-based semi-supervised mulit-label classification method achieves considerably higher performance in comparison with the performance of modified Xception net.
 
\section{Methodology}

The proposed Deep subspace analysis for semi-supervised multi-label classification is constructed of 3 main stages, detailed in bellow sections. Fig.~\ref{fig2} Demonstrates the summary of the proposed method.

\subsection{Stage.1: Transfer Learning} 

Number of recent semi-supervised feature extraction methods focused on finding latent representations of the input data using deep neural network. \\
Unlike those methods, we first employ the concept of transfer learning , on the labeled data only, to generate discriminative features. \\
That separates the features extraction technique from the rest of the classification stages, and results in massive time and cost reduction for retraining the method, when more unlabeled data become available. 

\begin{figure}
\includegraphics[width=\textwidth]{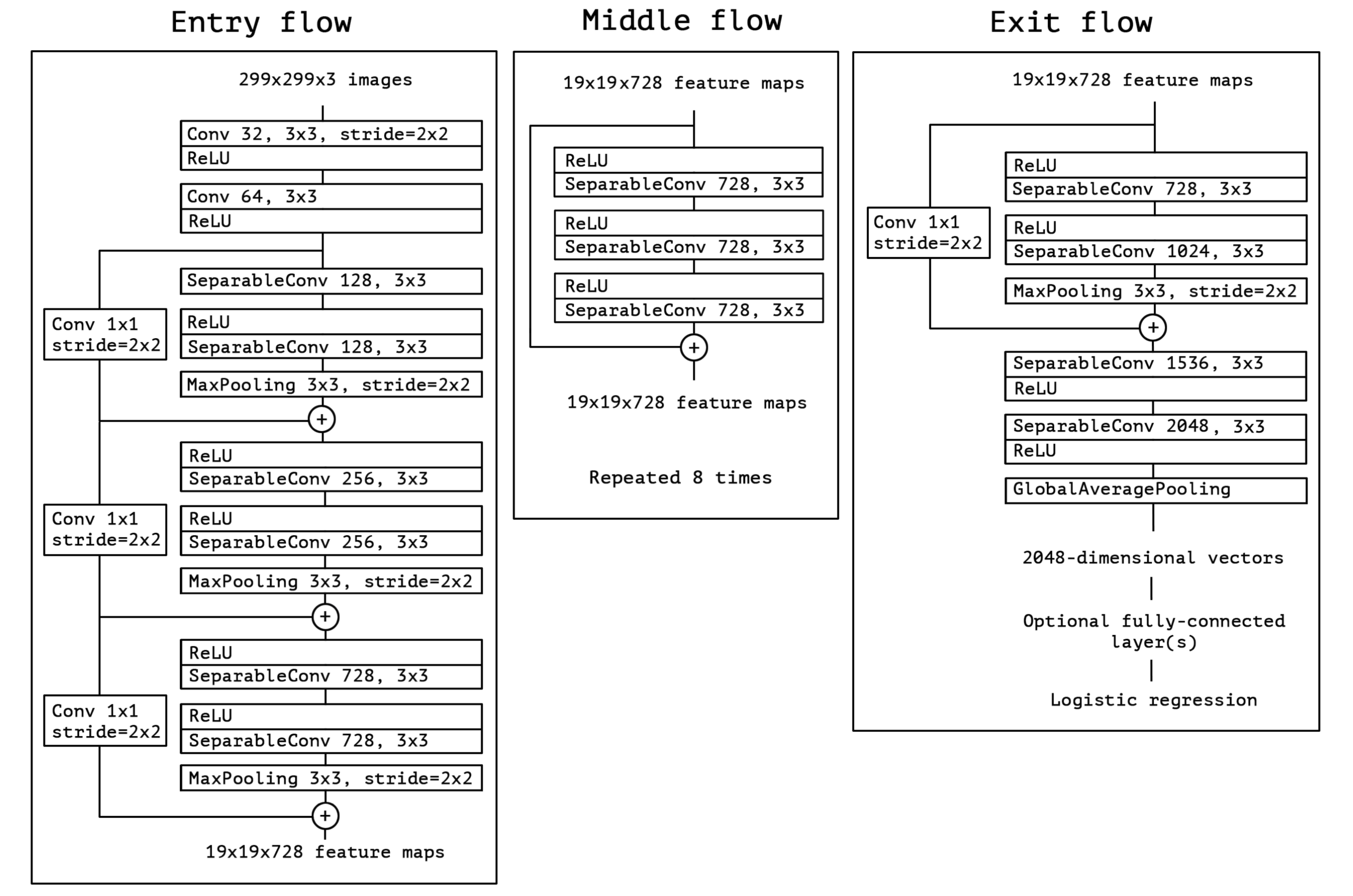}
\caption{The above image provides the detail of Xception architect \cite{Xception} } \label{fig1}
\end{figure}

\begin{figure} 
\begin{center}
\includegraphics[width=0.90\textwidth]{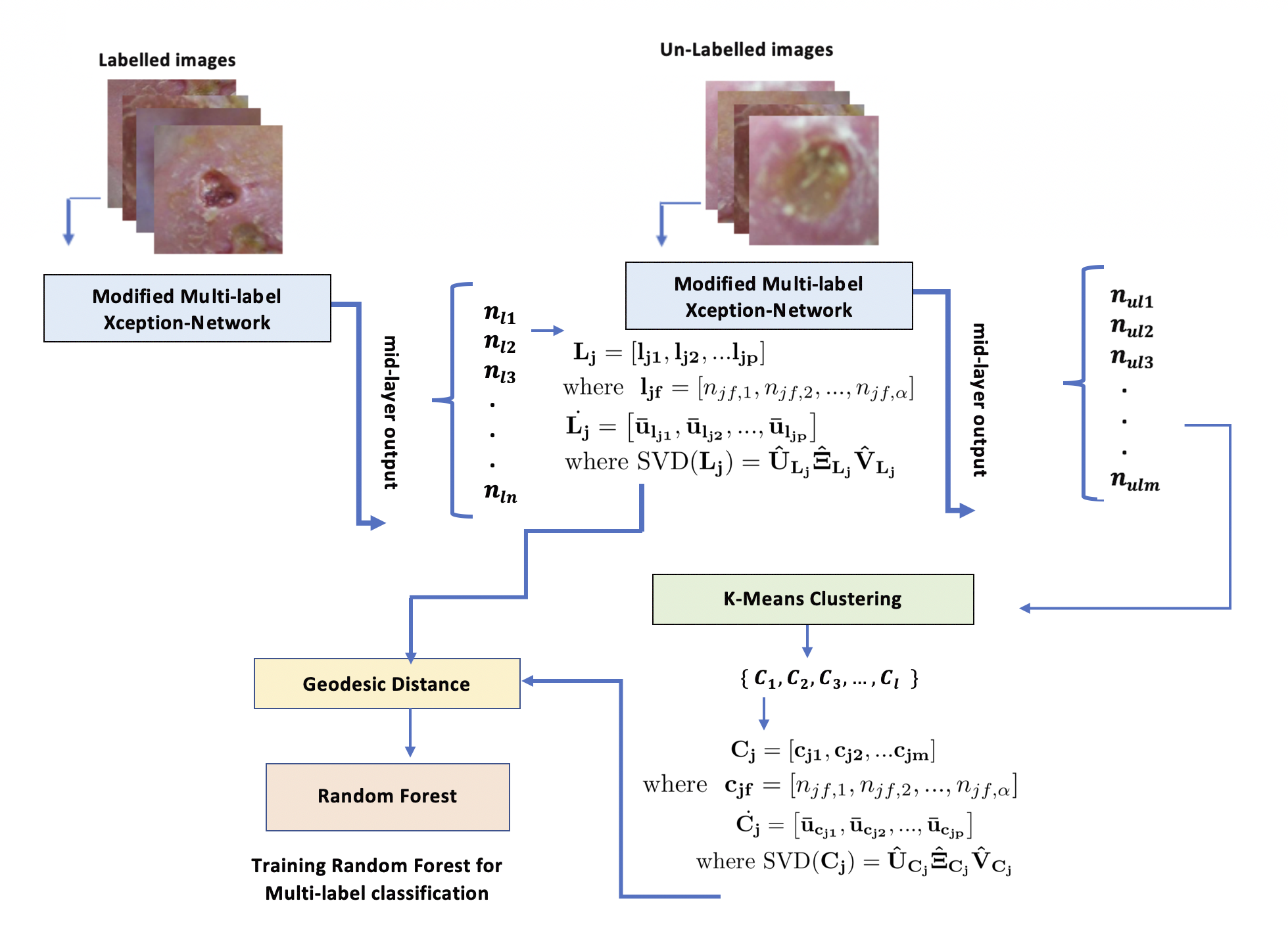}
\end{center}
\caption{The above image provides a summary of the proposed DSSC pipeline.} \label{fig2}
\end{figure}

We use Xception (Fig.~\ref{fig1}) with imagenet weights, as our base method. Extreme Inception (Xception) ~\cite{Xception}  is a deep convolutional neural network based architecture that have gained its success through decoupling the mapping of cross-channels correlations and spatial correlations in the feature maps of convolutional neural networks. In other words,  Xception network is a linear stack of depth-wise separable convolution layers with residual connections. The Xception network architect have following characteristics: 
\begin{itemize}
    \item That contains 36 convolutional layers. 
    \item The 36 convolutional layers are structured into 14 modules.
    \item All the modules, except the first and the last, have linear residual connections around them.
\end{itemize}

Specifically, in stage.1, we modify Xception network by removing the last layer and adding two fully connected layers of size 128 and 2 to re-train the network for multi-label classification.\\ To address the imbalanced data, we increase the penalty weight for Ischaemia class compared to the one for Infection class. 

Then, following the common practice, we train the modified Xception network in two steps; when first we freeze the original layers and train the final two layers. Then, we train the entire network with a lower training rate. 
Finally, the output of a mid-layer is extracted to represent the descriptive features of each given image.

\subsection{Stage.2: Deep Subspace-based descriptors} 

In this stage, to ensure that each image representation is shift invariant we employ image augmentation tools and generate an image set for each image. Then we map the image sets into points on Grassmann manifold by generating a linear subspace representation for each image. Finally, in order to take advantage of the unlabeled data, we use geodesic distance to compute the distance between labeled and number of unlabeled data; then, represent each image as a vector of relations. \\

In this section, we first explain the Grassmann Manifold topological space and geodesic distance that is used to calculate distance between point on Grassmann Manifold. Then we explain the transformation of an image into deep subspace-based relational vector.

\subsubsection{Grassmann Manifold}

In this study we are interested in two types of Riemannian manifolds, namely the Grassmann manifolds and
the manifolds of Symmetric, Positive Definite matrices (SPD).
Manifolds are smooth, curved surfaces embedded in higher dimensional Euclidean spaces and formally defined as follows:

\begin{definition}
A topological space {\small $\mathcal{M}$} is called a manifold if:
\begin{itemize}

  \item
  {\small $\mathcal{M}$} is Hausdorff
  \footnote
    {
    In a Hausdorff space, distinct points have disjoint neighbourhoods.
    This property is important to establish the notion of a differential manifold,
    as it guarantees that convergent sequences have a single limit point.
    }%
    ,
    i.e. every pair {\small $\mathbf{X}$}, {\small $\mathbf{Y}$} can be separated by two disjoint open sets.
    
 \item
  {\small $\mathbf{M}$} is locally Euclidean,
  that is, for every
  \text{\small $\mathbf{X} \in \mathbf{M}$}
  there exists an open set
  \text{\small $U \subset \mathbf{M}$} with \text{$\mathbf{X} \in \mathbf{U}$}
  and an open set
  \text{\small $V \subset \mathbf{R}^{n}$} with a homeomorphism \text{\small $\varphi:U \rightarrow V$}.
\end{itemize}
\end{definition}

To formally define a Grassmann manifold and its geometry,
we need to define the quotient space of a manifold.
A quotient space of a manifold, intuitively speaking,
is the result of ``gluing together'' certain points of the manifold.
Formally, given {\small $\sim_{\psi}$} as an equivalence relation on {\small $\mathbf{M}$},
the quotient space \mbox{\small $\Upsilon = \mathbf{M}/\sim_{\psi}$}
is defined to be the set of equivalence classes of elements of {\small $\mathbf{M}$},
i.e.
\mbox{\small $\Upsilon=\{[\mathbf{X}]:\mathbf{X} \in \mathbf{M}\}
= \{[\mathbf{Y} \in \mathbf{M}:\mathbf{Y} \sim_{\psi} \mathbf{X}]:\mathbf{X} \in \mathbf{M}\}$}.

\begin{definition}
A Grassmann manifold is a quotient space of the special orthogonal group%
\footnote{
Special orthogonal group {\small $SO(n)$} is the space of all \mbox{$n \times n$} orthogonal matrices with the  determinant $+1$.
It is not a vector space but a differentiable manifold, i.e., it can be locally approximated by subsets of a Euclidean
space.
}
{\small $SO(n)$} and is defined as a set of {\small $p$}-dimensional linear subspaces of~{\small $\mathbf{R}^n$}.
\end{definition}
In practice an element {\small $\mathbf{X} $} of {\small $\mathbf{G}{n,p}$} is represented by an orthonormal basis as a
{\small$n \times p $} matrix, i.e., {\small $\mathbf{X}^T \mathbf{X} = \mathbf{I}_p$}.
The geodesic distance between two points on the Grassmann manifold can be computed as:
\begin{equation}
    d_G\left(\mathbf{X},\mathbf{Y}\right)=\|\Theta\|_2
    \label{eqn:geodesic_Grass}
\end{equation}%

\noindent
where \mbox{$\Theta=[\theta_1,\theta_2,\cdots,\theta_p]$} is the principal angle vector,
i.e.:
\begin{equation}
  \cos(\theta_i)
  =
  \max_{\Vec{x}_i \in \mathbf{X}, ~ \Vec{y}_j \in \mathbf{Y}}
  \Vec{x}_i^T \Vec{y}_j
  \label{eqn:Principal_Angle}
\end{equation}%

\noindent
subject to
\mbox{\small $\mathbf{x}_i^T \mathbf{x}_i \text{~=~} \mathbf{y}_i^T \mathbf{y}_i \text{~=~} 1$},
\mbox{\small $\mathbf{x}_i^T \mathbf{x}_j \text{~=~} \mathbf{y}_i^T \mathbf{y}_j \text{~=~} 0$},
\mbox{\small $i \neq j$}.
The principal angles have the property of \mbox{$\theta_i \in [0, \pi/2]$}
and can be computed through SVD of \mbox{\small $\mathbf{X}^T \mathbf{Y}$}.

\subsubsection{Geodesic-based Relational Representation}

First, we employ K-medians clustering on unlabeled data to generate their representatives. Then we generate linear subspace for each labeled data and the centroid images of unlabeled data.

Next, to generate a linear subspace representative for each image, we start by employing data augmentation to generate image set representation of each image. Then, we represent each image as an output of the mid layer of modified Xception network; which is then followed by calculating the strongest Eigen vectors through computing Singular Value Decomposition (SVD) of the deep image-set. That maps each image set into a linear subspace which is a point on Grassmann Manifold.

Then, we calculate the geodesic distance between each labeled image to the K-medians centers of unlabeled images. That is constructed by calculating the distance between extracted features computed in stage.1.\\
Let $\breve{\mathbf{L}}$ represent the labeled training data, and $\breve{\mathbf{U}}$ represent the Un-labeled training data:

\begin{gather*} 
        \breve{\mathbf{L}} = \left [ \mathbf{l_1}, \mathbf{l_2},  ... \mathbf{l_m}\right ] \\
        \mathbf{l_i} = \left [ n_{i,1}, n_{i,2},  ... ,n_{i,128}\right ]
\\ \\
\breve{\mathbf{U}} = \left [ \mathbf{u_1}, \mathbf{u_2},  ... \mathbf{u_p}\right ] \\
\mathbf{u_j} = \left [ n_{j,1}, n_{j,2},  ... ,n_{j,128}\right ]
\end{gather*}
, where vector $\mathbf{l_i}$ and $\mathbf{u_j}$ are the deep representation of a labeled image $I_i$, and un-labled image $U_j$ respectively. That is the feature vector extracted from the mid-layer of the modified Xception architect.\\
After performing Kmedians on un-labeled training data, the un-labled data would be represented as matrix $\breve{\mathbf{C}}$:
\begin{gather*}
    \breve{\mathbf{C}} = \left [ \mathbf{c_1}, \mathbf{c_2},  ... \mathbf{c_p}\right ] \\
\mathbf{c_j} = \left [ n_{j,1}, n_{j,2},  ... ,n_{j,\alpha }\right ]
\end{gather*}
, where vector $\mathbf{c_k}$ is the deep representation of a  centroid image $\mathbf{u_k}$. For this work we have used $~~ \alpha = 200$ , that means the Kmedians would compute the index of 200 centroids for un-labeled training data.\\

At this point, we transfer each image into an image set using augmentation techniques:
\begin{gather*}
  \forall \mathbf{L_j} \in \breve{\mathbf{L}} : \mathbf{L_j} = \left [ \mathbf{l_{j1}}, \mathbf{l_{j2}},  ... \mathbf{l_{jp}}\right ] ~~
    \text{ where}~~
\mathbf{l_{jf}} = \left [ n_{jf,1}, n_{jf,2},  ... ,n_{jf,\alpha }\right ]\\ 
\dot{\mathbf{L_j}} = \left [ \mathbf{\bar{u}_{l_{j1}}}, \mathbf{\bar{u}_{l_{j2}}},  ... , \mathbf{\bar{u}_{l_{jp}}}\right ]~~
\text{ where SVD(} \mathbf{L_j} \text{)} = \mathbf{{\hat{U}}_{L_j}}  \mathbf{{\hat{\Xi}}_{L_j}}  \mathbf{{\hat{V}}_{L_j}}
\\ \text{, and}~~\\ 
  \forall \mathbf{C_j} \in \breve{\mathbf{C}} :  \mathbf{C_j} = \left [ \mathbf{c_{j1}}, \mathbf{c_{j2}},  ... \mathbf{c_{jm}}\right ] ~~
    \text{ where}~~
\mathbf{c_{jf}} = \left [ n_{jf,1}, n_{jf,2},  ... ,n_{jf,\alpha }\right ]\\ 
\dot{\mathbf{C_j}} = \left [ \mathbf{\bar{u}_{c_{j1}}}, \mathbf{\bar{u}_{c_{j2}}},  ... , \mathbf{\bar{u}_{c_{jp}}}\right ]~~
\text{ where SVD(} \mathbf{C_j} \text{)} = \mathbf{{\hat{U}}_{C_j}}  \mathbf{{\hat{\Xi}}_{C_j}}  \mathbf{{\hat{V}}_{C_j}}
\end{gather*}
, where $\dot{\mathbf{L_j}}$ and $\dot{\mathbf{C_j}}$ are each a linear subspace representing labeled image $L_j$, and a centroid of unlabeled data $C_j$ respectively.\\
Finally, we represent each Labeled image $\mathbf{l_i}$ through calculating the geodesic distance between $\dot{\mathbf{L_i}}$ and all the $\dot{\mathbf{C_j}} \in \dot{\breve{\mathbf{C}}}$. 
\begin{gather*}
   \text{for}~~\mathbf{L_j} \in \dot{\breve{\mathbf{L}}}~~\text{and,}~~ \mathbf{C_j} \in \dot{\breve{\mathbf{C}}}: \\
   \left \|  \mathbf{d_i} \right \| = \left ( G_d(\dot{\mathbf{L_j}} , \dot{\mathbf{C_1}}) ,  G_d(\dot{\mathbf{L_j}} , \dot{\mathbf{C_2}})  , ... ,  G_d(\dot{\mathbf{L_j}} , \dot{\mathbf{C_\alpha}})\right ) 
\end{gather*}
, where each image $\mathbf{l_i}$ is now represented as $\left \|  \mathbf{d_{Gi}} \right \|$ detailed in the equation. \ref{eqn:geodesic_Grass}.

\subsection{Stage.3: Final Classification}
Finally, we employ Multi-Label Random Forest (MLRF) \cite{liu2015mlrf} classification method on the resulting feature vector. 
MLRF, is a multi-label classification method based on a variation of random forest. It uses a new label set partition method to transform multi-label data sets into multiple single-label data sets. That can \ optimize the label subset partition, by discovering the correlated labels. That employs an on-line kNNs-like sampling method for each generated single-label subset ignorer to learn a random forest classifier .

\section{Results and Discussion}

To evaluate the proposed method, we have tested its performance on blind test set of Diabetic Foot Ulcers dataset (DFUC2021)\cite{DFUC2021}. We have used 5,955 DFU images for training, 5,734 for blind testing \cite{DFUC2021}. The ground truth labels comprise of four classes: control, infection, ischaemia and both conditions. The results indicate a considerable improvement when using the proposed semi-supervised method compared to solely relying on transfer learning through using the modified version of Xception. Bellow table summarises our finding.

It is important to note that more complex modification of Xception with more data augmentation would result in the better performance for both Xception and DSSC. That would be developed and evaluated in future work.

To generate bellow results, Xception was trained for 10 iterations in the first training step, and 40 for the second stage (with smaller learning rate). 

\begin{table}
\caption{AUC of proposed DSSC method validation on blind test set of DFU2021.}\label{tab1}
\begin{tabular}{|l|l|l|l|l|l|}
\hline
Method &  Both-AUC & None-AUC & Infection-AUC & Ischaemia-AUC & Macro-AUC \\
\hline
Modified Xception & 0.6483 & 0.7215 & 0.6438 &  0.7331 & 0.6867 \\
Proposed DSSC &  {\bfseries 0.7547} & {\bfseries 0.7443} &  {\bfseries 0.7024} &  {\bfseries 0.7382} &  {\bfseries 0.7349} \\
\hline
\end{tabular}
\end{table}

\begin{table}
\caption{F1-Score of proposed DSSC method validation on blind test set of DFU2021.}\label{tab2}
\begin{tabular}{|l|l|l|l|l|l|}
\hline
Method &  Both-F1 & None-F1 & Infection-F1 & Ischaemia-F1 & Macro-AF1 \\
\hline
Modified Xception & 0.3737 & 0.7112 & 0.5503 & 0.5111 & 0.5067 \\
Proposed DSSC &  {\bfseries  0.5314} & {\bfseries 0.7243} &  {\bfseries 0.6454} & {\bfseries 0.4708} &  {\bfseries 0.5930} \\
\hline
\end{tabular}
\end{table} 

In Table.1, and Table.2: Both- refers to where the image include both Infection and Ischaemia, None- refers to where image include no Infection or Ischaemia, and Infection- and Ischaemia- refers to where the image include only one condition respectfully. 

The above results indicate that the proposed DSSC method results in considerable improvement in performance compared to Xception, through taking advantage of unlabeled data, while ensuring the ease of retraining for new unlabeled data (when become available).

\section{Conclusion}
In this work, we propose a novel deep subspace analysis method for semi-supervised multi-label classification (DSSCC) of DFU images. The proposed method have two main differences compared to the recent state of the art deep semi-supervised methods. First, unlike recent research works in deep semi-supervised methods, the proposed pipeline dose not augment data during training; instead, to generate a shift invariant representative, it transfer each image into a linear subspace, and analysis them using Grassmann manifold geometry. Moreover, the method considered that the likelihood of new unlabelled data becoming available is always higher compared to that for labeled data. Thus, the proposed method is designed so the retraining of network with more unlabeled data would not be time and resource consuming. 
The evaluation of DSSC on blind test set of DFU2021 shows considerable improvement compared to the performance of solely relying on labeled data using Xception. That proves the efficiency of the proposed Deep Relation-based Semi-Supervised.

%
%
%
%
%
%

\end{document}